\documentclass[10pt]{article}
\usepackage{latexsym,graphicx}
\usepackage{latexsym,graphicx}
\newcommand{\be}{\begin{equation}}
\newcommand{\ee}{\end{equation}}
\def\n{\noindent}
\catcode `\@=11 \catcode `\@=12
\begin{document}
\begin{center}
\large{\bf {Bianchi Type V String Cosmological Models in General Relativity}} \\
\vspace{10mm}
\normalsize{ ANIL KUMAR YADAV\footnote{corresponding author}, VINEET
KUMAR YADAV$^2$, LALLAN YADAV$^3$ }\\
 \vspace{3mm}
$^1$\normalsize{Department of Physics, Anand Engineering
College, Keetham, Agra -282 007, India}\\
$^1$\normalsize{E-mail : abanilyadav@yahoo.co.in,~ akyadav@imsc.res.in}\\
\vspace{1mm}
$^{2,3}$\normalsize{Department of Physics, D. D. U. Gorakhpur
University, Gorakhpur -273 009, India}\\

$^2$\normalsize{E-mail : vineetky@yahoo.co.in}\\~~~~
$^3$\normalsize{E-mail : nisaly06@rediffmail.com}\\
\end{center}
\vspace{10mm}
\begin{abstract}
\n Bianchi type V string cosmological models in general relativity are investigated. 
To get the exact solution of Einstein's field equations, we have taken some scale transformations followed by 
Camci $\it et ~ al$ (2001). It is shown that Einstein's field equations are solvable for
any arbitrary cosmic scale function. Solution for particular form of cosmic scale functions are also obtained. 
Some physical and geometrical aspects of the models are discussed.\\
\end{abstract}
\smallskip
\n PACS: 98.80.cq, 98.80.-k\\
\n Key words: Cosmology, Bianchi type V universe, string theory.  
\section{Introduction}
The study of Bianchi type V cosmological models play an important
role in the study of universe and it create more interest as these
models contains isotropic special cases and permit arbitrary small
anisotropy levels at some instant of time.
The string theory plays a significant role in the study of physical situation at the very early
stages of the formation of the universe. It is generally assumed that after the big bang , 
the universe may have undergone a series of phase transitions as its temperature lowered down below some 
critical temperature as predicted by grand unified theories [1-6]. At the very early stages of evolution  
of the universe, it is believed that during phase transition the symmetry of the universe is broken spontaneously. 
It can give rise to topologically stable defects such as domain walls, strings and monopoles. 
In all these three cosmological structures, only cosmic strings have excited the most interesting consequence [7] 
because they are believed to give rise to density perturbations which lead to formation of galaxies[8]. 
These cosmic strings can be closed like loops or open like a hair which move through time and trace out a tube 
or a sheet, according to whether it is closed or open. The string is free to vibrate and its 
different vibrational modes present different types of particles carrying the force of gravitation. 
This is why much interesting to study the gravitational effect that arises from strings by using Einstein's 
field equations.

\vspace{1mm}

The general relativistic treatment of strings has been initially given by Stachel [9] and Letelier [10,11].
Letelier [10] obtained the general solution of Einstein's field equations for a cloud of strings with spherical,
plane and a particular case of cylindrical symmetry. Letelier [11] also obtained massive string
cosmological models in Bianchi type-I and Kantowski-Sachs space-times. Benerjee
et al. [12] have investigated an axially symmetric Bianchi type I string dust
cosmological model in presence and absence of magnetic field. Exact solutions of
string cosmology for Bianchi type II, $VI_{0}$, VIII and IX space-times have
been studied by Krori et al. [13] and Wang [14,15]. Bali and Upadhaya [16] have presented LRS Bianchi 
type I string dust magnetized  cosmological models. Singh  and Singh [17] investigated string cosmological 
models with magnetic field in the context of space-time with $G_{3}$ symmetry. Singh [18,19] has studied 
string cosmology with electromagnetic fields in Bianchi type II, VIII and IX space-times.

\vspace{1mm} 

Bianchi V universes are the natural generalization of FRW models with negative curvature.These open models 
are favored by the available evidences for low density universes (Gott et al [20]). Bianchi type V cosmological model 
where matter moves orthogonally to the hyper-surface of homogeneity, has been studied by Heckmann and Schucking [21].
Exact tilted solutions for the Bianchi type V space-time are obtained by Hawkings [22], Grishchuk et al. [23]. Ftaclas
and Cohen [24] have investigated LRS Bianchi type V universes containing stiff matter with electromagnetic field.
Lorentz [25] has investigated LRS Bianchi type V tilted models with stiff fluid and electromagnetic field. 
Pradhan et al have investigated The generation of Bianchi type V cosmological models with varying $\Lambda$ 
term. Yadav et al.[27,28] have investigated bulk viscous string cosmological models in 
different space-times. Bali and Anjali [29], Bali [30] have obtained 
Bianchi type-I, and type V string cosmological models in general 
relativity. The string cosmological models with a magnetic field are discussed by Tikekar and Patel [31],
Patel and Maharaj [32]. Ram and Singh [33] obtained some new exact solution of string 
cosmology with and without a source free magnetic field for Bianchi type I space-time in the different basic form 
considered by Carminati and McIntosh [34]. Yavuz et al. [35] have examined 
charged strange quark matter attached to the string cloud in the spherical symmetric 
space-time admitting one-parameter group of conformal motion. Kaluza-Klein 
cosmological solutions are obtained by Yilmaz [36] for quark matter attached to the 
string cloud in the context of general relativity. Recently, Baysal et al. [37], Kilinc and Yavuz [38],
 Pradhan [39] and Pradhan et al.[40,41], Yadav et al [42] have investigated some string cosmological 
models in cylindrically symmetric inhomogeneous universe. In this paper, we have investigated string cosmological 
models based on generation technique and obtained a more realistic behavior of the universe. This paper is organised
as follows. The field equations are presented in Section 2. In Section 3, we deal with generation technique for solution 
of field equation and finally the results are discussed in Section 4.  
\section{Field Equations}
We consider the Bianchi type V metric of the form
\begin{equation}
ds^2=dt^2-A^2(t)dx^2-e^{2\alpha x} [B^2(t)dy^2+C^2(t)dz^2]
\end{equation}
where $\alpha$ is a constant.\\
 The energy momentum tensor for cloud
of string is given by
\begin{equation}
T^{j}_{i} =\rho v_{i}v^{j} - \lambda x_{i}x^{j}
\end{equation}
where $v_{i}$ and $x_{i}$ satisfy the condition
\begin{equation}
v^{i}v_{j} = -x^{i}x_{j} = 1,  v^{i}x_{i} = 0.
\end{equation}
Here $\rho$ is the proper energy density of the cloud of string with
particle attached to them.$\lambda$ is the string tension density,
$v^{i}$, the four velocity of the particles and $x^{i}$, the
unit space vector representing the direction of strings.If the
particle density of the configuration is denoted by $\rho_{p}$ then
we have
\begin{equation}
\rho = \rho_{p} +\lambda
\end{equation}
For the energy momentum tensor (2) and Bianchi type V metric (1),
Einstien's field equations
\begin{equation}
R^{j}_{i}-\frac{1}{2}Rg_{ij} = -8\pi T_{ij}
\end{equation}
yield the following five independent equations
\vspace{3mm}
\begin{equation}
\frac{\ddot{A}}{A}+\frac{\ddot{B}}{B}+\frac{\dot{A}\dot{B}}{AB}-\frac{\alpha^2}{A^2}=0
\end{equation}
\begin{equation}
\frac{\ddot{A}}{A}+\frac{\ddot{C}}{C}+\frac{\dot{A}\dot{C}}{AC}-\frac{\alpha^2}{A^2}=0
\end{equation}
\begin{equation}
\frac{\ddot{B}}{B}+\frac{\ddot{C}}{C}+\frac{\dot{B}\dot{C}}{BC}-\frac{\alpha^2}{A^2}=-8\pi
\lambda
\end{equation}
\begin{equation}
\frac{\dot{A}\dot{B}}{AB}+\frac{\dot{A}\dot{C}}{AC}+\frac{\dot{B}\dot{C}}{BC}-\frac{3\alpha^2}{A^2}=-8\pi
\rho
\end{equation}
\begin{equation}
\frac{2\dot{A}}{A}-\frac{\dot{B}}{B}-\frac{\dot{C}}{C}=0
\end{equation}
Here and what follows the dots overhead the symbol A, B, C denotes
differentiation with respect to t.\\
The physical quantities expansion scalar $\theta$ and shear scalar $\sigma^2$ have the following expressions
\begin{equation}
 \theta = \frac{\dot{A}}{A}+\frac{\dot{B}}{B}+\frac{\dot{C}}{C}
\end{equation}
\vspace{2mm}
\begin{equation}
\sigma^2 =\frac{1}{2}\sigma_{ij}\sigma^{ij}=\frac{1}{3}\left[\theta^2-\frac{\dot{A}\dot{B}}{AB} 
-\frac{\dot{A}\dot{C}}{AC}-\frac{\dot{B}\dot{C}}{BC}\right]
\end{equation}
 
Integrating equation $(10)$ and absorbing the integrating constant into B or C, we obtain
\begin{equation}
 A^2=BC
\end{equation}
without loss of any generality.\\
From equations $(6)$, $(7)$ and $(13)$, we obtain
\begin{equation}
 2\frac{\ddot{B}}{B}+\left(\frac{\dot{B}}{B}\right
)^2 = 2\frac{\ddot{C}}{C}+\left(\frac{\dot{C}}{C}\right)^2
\end{equation}
which on integration yields
\begin{equation}
 \frac{\dot{B}}{B}-\frac{\dot{C}}{C} = \frac{k}{(BC)^{\frac{3}{2}}}
\end{equation}
where k is the constant of integration. Hence for the metric function B or C from the above first order differential
equation $(15)$, some scale transformations permit us to obtain new metric function B or C.\\
~~~ Firstly, under the scale transformation $dt=B^{\frac{1}{2}}dT$, equation $(15)$ takes the form
\begin{equation}
CB_{T}-BC_{T} = kC^{-\frac{1}{2}}
\end{equation}
where the subscript represents derivative with respect to $T$. Considering equation $(16)$ as a linear
differential equation for B, where C is an arbitrary function, we obtain
\begin{equation}
(i)~~~~~~~~~~~~~~~~~~~~B = k_{1}C+kC \int{\frac{dT}{C^{\frac{5}{2}}}},
\end{equation}
where $k_{1}$ is the the constant of integration. Similarly, using the transformation $dt=B^\frac{3}{2}d\bar{T}$,
$dt=C^\frac{1}{2}d\tilde{t}$ and $dt=C^\frac{3}{2}d\tau$ in equation $(15)$
after some algebra we obtain respectively.
\begin{equation}
(ii)~~~~~~~~~~~~~~~~~~~~~~B(\bar{T})=k_{2}Ce^{\left(k\int\frac{d\bar{T}}{C^\frac{3}{2}}\right)},
\end{equation}
\begin{equation}
(iii)~~~~~~~~~~~~~~~~~~~~~~C(\tilde{t})=k_{3}B-kB\int\frac{d\tilde{t}}{B^\frac{5}{2}},
\end{equation}
\begin{equation}
(iv)~~~~~~~~~~~~~~~~~~~~~~~C(\tau)=k_{4}Be^{\left(k\int\frac{d\tau}{B^\frac{3}{2}}\right)},
\end{equation}
where $k_{2}$, $k_{3}$ and $k_{4}$ are constant of integration. Thus choosing any given function B or C in 
cases (i), (ii), (iii) and (iv), one can obtain B or C.\\
\section{Generation technique for solution}
We consider the following four cases
\subsection{Case (i):~$ C=T^n $~~(n~is~a~real~number~satisfying~$ n\neq\frac{2}{5}$)}
Equation $(16)$ leads to
\begin{equation}
 B=k_{1}T^n+\frac{2k}{2-5n}T^{1-\frac{3n}{2}}
\end{equation}
From equations $(13)$ and $(21)$, we obtain
\begin{equation}
 A^2=k_{1}T^{2n}+\frac{2k}{2-5n}T^{1-\frac{n}{2}}
\end{equation}
Hence metric $(1)$ reduces to the following form
\begin{equation}
 ds^2=(k_{1}T^n+2lT^{l_{1}})[dT^2-T^ndx^2]-e^{2\alpha x}\left[\left(k_{1}T^n+2lT^{l_{1}}\right)^2dy^2+T^{2n}dz^2\right],
\end{equation}
where $l=\frac{k}{2-5n}$ and $l_{1}=1-\frac{3n}{2}$\\
In this case the physical parameters, i.e. the string tension density $(\lambda)$, the energy density $(\rho)$, 
the particle density $(\rho_{p})$ and the kinematical parameters, i.e. the scalar of expansion $(\theta)$,
shear scalar $(\sigma)$ and the proper volume $(V^3)$ for model $(23)$ are given by\\
\[
8\pi\lambda=\left[-2k_{1}^2n(n-1)T^{2n-2}-k_{1}ln(10-13n){T^{-(l_{1}+2n)}}-\frac{1}{2}l^2(4+4n-11n^2){T}^{-3n}\right]
\]
\begin{equation}
 \left(k_{1}T^n+2lT^{l_{1}}\right)^{-3}+\alpha^2{T}^{-n}\left(k_{1}T^n+2lT^{l_{1}}\right)^{-1},
\end{equation}
\vspace{2mm}
\[
8\pi\rho=\left[3k_{1}^2n^2T^{2n-2}+3k_{1}ln(2-n)T^{-(l_{1}+2n)}+\frac{1}{2}l^2(4+4n-11n^2)T^{-3n}\right]
\]
\begin{equation}
\left(k_{1}T^n+2lT^{l_{1}}\right)^{-3}-3\alpha^2T^{-n}\left(k_{1}T^n+2lT^{l_{1}}\right)^{-1},
\end{equation}
\vspace{2mm}
\[
8\pi\rho_{p}=\left[n(5n-2)k_{1}^2T^{2n-2}+16k_{1}ln(1-n)T^{-(l_{1}+2n)}+l^2(4+4n-11n^2)T^{-3n}\right]
\]
\begin{equation}
\left(k_{1}T^n+2lT^{l_{1}}\right)^{-3}-4\alpha^2T^n\left(k_{1}T^n+2lT^{l_{1}}\right)^{-1}
\end{equation}

\vspace{2mm}
\begin{equation}
 \theta=3\left[k_{1}nT^{n-1}+\frac{1}{2}l(2-n)T^{\frac{-3n}{2}}\right]\left(k_{1}T^n+2lT^{l_{1}}\right)^{-\frac{3}{2}}
\end{equation}
\vspace{2mm}
\begin{equation}
 \sigma=\frac{1}{2}kT^{-\frac{3n}{2}}\left(k_{1}T^{n}+2lT^{l_{1}}\right)^{-\frac{3}{2}}~~~~~~~~~~~~~~~~~~~~~~~~~~~~~~
\end{equation}
\vspace{2mm}
\begin{equation}
 V^3=\left(k_{1}T^{2n}+2lT^{n+l_{1}}\right)^\frac{3}{2}e^{2\alpha x}~~~~~~~~~~~~~~~~~~~~~~~~~~~~~~
\end{equation}
Equations $(27)$ and $(28)$ leads to
\vspace{2mm}
\begin{equation}
\frac{\sigma}{\theta}=\frac{1}{6}k\left[k_{1}nT^{n-l_{1}} +\frac{1}{2}l(2-n)\right]^{-1}~~~~~~~~~~~~~~~~~~~~~~~
\end{equation}
The energy condition $ \rho\geq0 $ and $ \rho_{p}\geq0 $ satisfy for model $(23)$. The condition $ \rho\geq0 $ and
$ \rho_{p}\geq0 $ leads to\\
\[
\left[3k_{1}^{2}n^{2}T^{3n-2}+3k_{1}ln(2-n)T^{-(l_{1}+n)}+\frac{1}{2}l^2
\left(4+4n-11n^2\right)T^{-2n}\right]
\]
\begin{equation}
\left(k_{1}T^n+2lT^{l_{1}}\right)^{-2}\geq{3\alpha}^2
\end{equation}
\vspace{2mm}
\[
 \left[n(2-5n)k_{1}^{2}T^{3n-2}+16k_{1}ln(n-1)T^{-(l_{1}+n)}+l^2\left(11n^2-4n-4\right)T^{-2n}\right]
\]
\begin{equation}
\left(k_{1}T^n+2lT^{l_{1}}\right)^{-2}\geq{4\alpha}^2
\end{equation}
respectively.\\
We observe that the string tension density $\lambda\geq0$, leads to
\vspace{2mm}
\[
\left[-2k_{1}^{2}n(n-1)T^{3n-2}-k_{1}ln(10-13n)T^{-(l_{1}+n)}-\frac{1}{2}l^2
\left(4+4n-11n^2\right)T^{-2n}\right]
\]
\begin{equation}
\left(k_{1}T^n+2 lT^{l_{1}}\right)^{-2}\geq{-\alpha}^2
\end{equation}
Generally the model $(23)$ are expanding, shearing and approaches to isotropy at late time. For $k=0$, 
the solution represents the shear free model of universe. We observe that as $T\rightarrow\infty$, $V^3\rightarrow\infty$ and $\rho\rightarrow0$
hence volume increases when T increases and the proper energy density of the cloud of string with particle 
attached to them decreases i. e. the proper energy density $(\rho)$ is decreasing function of time.\\ 
\subsection{case(ii): $ C={\bar{T}}^{n} $~~(n is a real number satisfying  $n\neq\frac{2}{3}$)}
Equation $(18)$ leads to 
\begin{equation}
 B=k_{2}{\bar{T}}^{n}e^{M{\bar{T}}^{l_{1}}}
\end{equation}
From equations $(13)$, and $(34)$, we obtain
\begin{equation}
 A^{2}=k_{2}{\bar{T}}^{2n}e^{M{\bar{T}}^{l_{1}}}
\end{equation}
where $M=\frac{k}{l_{1}}$.
Hence the metric $(1)$ reduces to the form
\begin{equation}
 ds^2={\bar{T}}^{\frac{4(1-l_{1})}{3}}\left[{\bar{T}}^{\frac{2(1-l_{1})}{3}}e^{3M{\bar{T}}^{l_{1}}}d{\bar{T}^2}
-e^{M{\bar{T}}^{l_{1}}}dx^2-e^{2\alpha x}\left(e^{2M{\bar{T}}^{l_{1}}}dy^2+dz^2\right)\right],
\end{equation}
Here $k_{2}$ is constant and we have taken $k_{2}$ as unity with out any loss of generality.
In this case the physical parameters, i.e. the string tension density $(\lambda)$, the energy density $(\rho)$, 
the particle density $(\rho_{p})$ and the kinematical parameters, i.e. the scalar of expansion $(\theta)$,
shear scalar $(\sigma)$ and the proper volume $(V^3)$ for model $(36)$ are given by\\
\begin{equation}
 8\pi\lambda=2n{\bar{T}}^{2(l_{1}-2)}+3nk{\bar{T}}^{3l_{1}-4}+\frac{1}{2}k^2{\bar{T}}^{4(l_{1}-1)}+
\alpha^2{\bar{T}}^{\frac{4(l_{1}-1)}{3}}e^{-3M\bar{T}^{l_{1}}},
\end{equation}
\vspace{2mm}
\begin{equation}
 8\pi\rho=3n^2{\bar{T}}^{2(l_{1}-2)}+3nk{\bar{T}}^{3l_{1}-4}+\frac{1}{2}k^2{\bar{T}}^{4(l_{1}-1)}-
3\alpha^2{\bar{T}}^{\frac{4(l_{1}-1)}{3}}e^{-3M\bar{T}^{l_{1}}},
\end{equation}
\vspace{2mm}
\begin{equation}
 8\pi\rho_{p}=n(3n-2){\bar{T}}^{2(l_{1}-2)}-4\alpha^2{\bar{T}}^{\frac{4(l_{1}-1)}{3}}e^{3M\bar{\bar{T}}^{l_{1}}}~~~~~~~~~~~~~~~~~~~~~~~~~~~~~~~~~~~~~~~
\end{equation}
\begin{equation}
 \theta=3\left[n{\bar{T}}^{l_{1}-2}+\frac{1}{2}k{\bar{T}}^{2(l_{1}-1)}\right],
\end{equation}
\vspace{2mm}
\begin{equation}
\sigma=\frac{1}{2}k\bar{T}^{2(l_{1}-1)}e^{-3M\bar{T}^{l_{1}}} ~~~~~~~~~~
\end{equation}
\begin{equation}
 V^{3}=\left(k_{3}\bar{T}^{2n}e^{{M}\bar{T}^{l_{1}}}\right)^{\frac{3}{2}}e^{2\alpha x}~~~~~~~~~~~~
\end{equation}
From equation$(40)$ and $(41)$ leads to
\begin{equation}
 \frac{\sigma}{\theta}=\frac{k}{6(n\bar{T}^{-l_{1}}+\frac{1}{2}k)}~~~~~~~~~~~~~~~~
\end{equation}
The energy condition $ \rho\geq0 $ and $ \rho_{p}\geq0 $ are satisfy for model $(36)$. The condition $ \rho\geq0 $ and
$ \rho_{p}\geq0 $ leads to\\
\begin{equation}
e^{3M\bar{T}^{l_{1}}}\left[3n^2\bar{T}^{\frac{2l_{1}-8}{3}}+3nk\bar{T}^{\frac{5l_{1}-8}{3}}+
\frac{1}{2}k^2\bar{T}^{\frac{8l_{1}-8}{3}}\right]\geq3\alpha^2~~~~~~~~~~~~~~~~~~~~~~~~
\end{equation}
\vspace{2mm}
\begin{equation}
 n(3n-2)e^{3M\bar{T}^{l_{1}}}\bar{T}^{\frac{(2l_{1}-8)}{3}}\geq4\alpha^2 ~~~~~~~~~~~~~~~~~~~~~~~~~~~~~~~~~~
\end{equation}
respectively.\\
we observe that the string tension density $\lambda\geq0$, leads to
\vspace{2mm}
\begin{equation}
e^{3M\bar{T}^{l_{1}}}\left[2n\bar{T}^{\frac{2l_{1}-8}{3}} + 3nk\bar{T}^{\frac{5l_{1}-8}{3}}+
\frac{1}{2}k^2\bar{T}^{\frac{8l_{1}-8}{3}}\right]\geq-\alpha^2 
\end{equation}
For $l_{1}>2$, model $(36)$ is expanding and for $ l_{1}<2 $, model starts with big bang singularity.
Generally the model is expanding, shearing and approaches to isotropy at late time. For $k=0$, 
the solution represents the shear free model of universe. We observe that as $\bar{T}\rightarrow\infty$, $V^3\rightarrow\infty$ and $\rho\rightarrow0$
hence volume increases when $\bar{T}$ increases and the proper energy density of the cloud of string with particle 
attached to them decreases i. e. the proper energy density $(\rho)$ is decreasing function of time.\\ 
\subsection{Case(iii): $ B={\tilde{t}}^n $~~(n is a real number)}
Equation $(19)$ leads to
\begin{equation}
C=k_{3}{\tilde{t}^n}-2l{\tilde{t}}^{l_{1}}
\end{equation}
From Equations $(13)$ and $(47)$, we obtain
\begin{equation}
A^2=k_{3}{\tilde{t}}^{2n}-2l{\tilde{t}}^{l_{1}+n}
\end{equation}
Thus the metric $(1)$ reduces to the form
\begin{equation}
ds^2=\left(k_{3}{\tilde{t}}^n-2l{\tilde{t}}^{l_{1}}\right)\left[dt^2-{\tilde{t}}^{n}dx^2\right]-e^{2\alpha x}\left[
{\tilde{t}}^{2n}dy^2+\left(k_{3}{\tilde{t}}^{n}-2l{\tilde{t}}^{l_{1}}\right)^{2}dz^2\right],
\end{equation}
In this case the physical parameters, i.e. the string tension density $(\lambda)$, the energy density $(\rho)$, 
the particle density $(\rho_{p})$ and the kinematical parameters, i.e. the scalar of expansion $(\theta)$,
shear scalar $(\sigma)$ and the proper volume $(V^3)$ for model $(49)$ are given by\\
\[
 8\pi\lambda=\left[-\frac{1}{2}l^{2}(11n^2-4n-4){\tilde{t}}^{-3n}+lk_{3}n(13n-10){\tilde{t}}^{l_{1}+n}
-2k_{3}^2n(n-1){\tilde{t}}^{2n-2}\right]
\]
\begin{equation}
\left(k_{3}{\tilde{t}}^n-2l{\tilde{t}}^{l_{1}}\right)^{-3}+\alpha^2{\tilde{t}}^{-n}\left(k_{3}{\tilde{t}}^n-2l{\tilde{t}}^{l_{1}}\right)^{-1},
\end{equation}
\vspace{2mm}
\[
 8\pi\rho=\left[-\frac{1}{2}l^{2}(11n^2-4n-4){\tilde{t}}^{-3n}-3lk_{3}n(2-n){\tilde{t}}^{l_{1}+n}
+3k_{3}^{2}n^2{\tilde{t}}^{2n-2}\right]~~~~~~~~
\]
\begin{equation}
\left(k_{3}{\tilde{t}}^n-2l{\tilde{t}}^{l_{1}}\right)^{-3}-3\alpha^2{\tilde{t}}^{-n}\left(k_{3}{\tilde{t}}^n-2l{\tilde{t}}^{l_{1}}\right)^{-1}
\end{equation}
\vspace{2mm}
\[
 8\pi\rho_{p}=\left(k_{3}\tilde{t}^n-2l\tilde{t}^{l_{1}}\right)^{-3}\left[
  n(5n-2)k_{3}^2\tilde{t}^{2n-2}-lk_{3}n(12n-8)\tilde{t}^{l_{1}+n}\right]~~~~~~~~~~~~~~
 \]
\begin{equation}
-4\alpha^2\tilde{t}^n\left(k_{3}\tilde{t}^n
-2l\tilde{t}^{l_{1}}\right)^{-1}~~~~~~~~~
\end{equation}
\begin{equation}
\theta=3\left[\frac{1}{2}l(n-2){\tilde{t}}^{-\frac{3n}{2}}+k_{3}n\tilde{t}^{n-1}\right]
\left(k_{3}{\tilde{t}}^n-2l{\tilde{t}}^{l_{1}}\right)^{-\frac{3}{2}},
\end{equation}
\vspace{2mm}
\begin{equation}
\sigma=\frac{1}{2}k{\tilde{t}}^{\frac{-3n}{2}}\left(k_{3}{\tilde{t}}^{n}-2l\tilde{t}^{l_{1}}
\right)^{-\frac{3}{2}}~~~~~~~~~~~~~~~~~~~~~~~~~~~~
\end{equation}
\begin{equation}
V^3=\left(k_{3}{\tilde{t}}^{2n}-2l{\tilde{t}}^{l_{1}+n}\right)^{\frac{3}{2}}e^{2\alpha x}~~~~~~~~~~~~~~~~~~~~~~~~~~~~
\end{equation}
Equation$(53)$ and $(54)$ leads to
\begin{equation}
\frac{\sigma}{\theta}=\frac{k}{6}\left[k_{3}n{\tilde{t}}^{-(l_{1}+n)}+\frac{l(n-2)}{2}\right]^{-1}
\end{equation}
The energy condition $\rho\geq0$ and $\rho_{p}\geq0$ are satisfy for model $(49)$. The condition $\rho\geq0$ and
$ \rho_{p}\geq0 $ leads to\\
\[
\left[-\frac{1}{2}l^2\left(11n^2-4n-4\right)\tilde{t}^{-2n}-3lk_{3}n(2-n)\tilde{t}^{l_{1}+2n}+3k_{3}^2n^2
\tilde{t}^{(3n-2)}\right]
 \]
\begin{equation}
\left(k_{3}\tilde{t}^n-2l\tilde{t}^{l_{1}}\right)^{-2}\geq3\alpha^2
\end{equation}
\vspace{2mm}
\begin{equation}
\left(k_{3}\tilde{t}^n-2l\tilde{t}^{l_{1}}\right)^{-2}\left[n(5n-2)k_{3}^2\tilde{t}^{n-2}-lk_{3}n(12n-8)\tilde{t}^
l_{1}\right]\geq4\alpha^2
\end{equation}
respectively.\\
We observe that the string tension density $\lambda\geq0$, leads to
\vspace{2mm}
\[
 \left[-\frac{1}{2}l^2\left(11n^2-4n-4\right)\tilde{t}^{-2n}+lk_{3}(13n-10)\tilde{t}^{l_{1}+2n}-2k_{3}^2n(n-1)
 \tilde{t}^{3n-2}\right] 
\]
\begin{equation}
\left(k_{3}\tilde{t}^n-2l\tilde{t}^{l_{1}}\right)^{-2}\geq-\alpha^2
\end{equation}
Thus we see that the model $(49)$ is generally expanding, shearing and approaches to isotropy at late time.
For $k=0$, the solution represents the shear free model of universe. We observe that as $\tilde{t}\rightarrow\infty$, $V^3\rightarrow\infty$ and $\rho\rightarrow0$
hence volume increases when $\tilde{t}$ increases and the proper energy density of the cloud of string with particle 
attached to them  decreases i. e. the proper energy density $(\rho)$ is decreasing function of time.\\ 

\subsection{Case(iv): $ B=\tau^n $~~ (n is any real number)}
Equation $(20)$ leads to
\begin{equation}
C=k_{4}\tau^ne^{\left(\frac{k}{l_{1}}\tau^{l_{1}}\right)}
\end{equation}
From equation $(13)$ and $(60)$, we obtain
\begin{equation}
 A^2=k_{4}\tau^{2n}e^{\left(\frac{k}{l_{1}}\tau^{l_{1}}\right)}
\end{equation}
Hence the metric $(1)$ reduces to
\begin{equation}
ds^2=\tau^{2n}e^{\left(\frac{k}{l_{1}}\tau^{l_{1}}\right)}\left[\tau^ne^{\left(\frac{2k}{l_{1}}\tau^{l_{1}}\right)}d\tau^2
-dx^2\right]-e^{2\alpha x}\left[dy^2+e^{\left(\frac{2k}{l_{1}}\tau^{l_{1}}\right)}dz^2\right],
\end{equation}
Here $k_{4}$ is constant and we have taken $k_{4}$ as unity without any loss of generality. In this case we see that 
A, B, and C are exponential function as that of in  case $(ii)$. Thus the physical and geometrical 
properties of the model $(62)$ are similar to model $(36)$.\\
\section{Conclusion}
If we choose $\alpha=0$, metric $(1)$ becomes Bianchi type I metric, studied by several authors in different context.
In this paper, we have applied the technique followed by Camci et al $[43]$ for solving Einstein's field equations 
and found new solution for string
cosmological model. It is shown that the Einstein's field equation are solvable for an arbitrary cosmic scale function.
Starting from a particular cosmic function, new classes of spatially homogeneous and anisotropic cosmological models have been 
investigated for which the string fluid are rotation free but they do have expansion and shear.
It is also observed that in all cases, the physical and geometrical behavior of models are similar. 
Generally the models are expanding, shearing and non rotating. All the models are isotropized at late time.\\

In case $(i)$, for $n\geq1$, Model $(23)$ start expanding with big bang singularity 
and for $n\leq0$, Bianchi type-V universe preserve expanding nature as $T\rightarrow0$,~$\theta\rightarrow0$ and 
$T\rightarrow\infty$,~$\theta\rightarrow\infty$. It is also observed that for $k=0$, shear scalar $(\sigma)$ vanishes 
and model becomes isotropic.
In case $(ii)$, we observed that for $ l_{1}<2$, model $(36)$ started with big bang singularity
and expand through out the evolution of universe and for $l_{1}>2$, model $(36)$ also preserve the 
expanding nature as $\bar{T}\rightarrow0$,~$\theta\rightarrow0$ and $\bar{T}\rightarrow\infty$,~
$\theta\rightarrow\infty$. From equation $(41)$, it is clear that when $k=0$, the shear scalar $(\sigma)$ vanishes and
model becomes isotropic. In case $(iii)$, it is observed that model $(49)$ started with big bang singularity 
and expanding through the evolution of universe. From equation $(54)$, it is clear that for $k=0$, 
the shear scalar $(\sigma)$ vanishes and model isotropizes. In case $(iv)$, it is observed that the physical 
and geometrical properties of the model $(62)$ are similar to model $(36)$ , i.e the case $(ii)$. We note that in all 
cases, $\rho(t)$ is a decreasing function of time and it is alway positive. Further it is observed that for 
sufficiently large time, $\rho_{p}$ and $\lambda$ tend to zero. Therefore the strings disappear from the universe 
at late time (i. e. present epoch). The same is predicted by current observations. \\

\n {\large{\bf{Acknowledgements}}}

\vspace{1mm}

\n The authors are extremely grateful to the referee for his fruitful comments. Author (AKY) is thankful to IMSc, Chennai 
and HRI, Allahabad for providing research facilities .

\vspace{3mm}


\end{document}